\documentclass{phb-proc4-auth}

\usepackage{graphicx}
\usepackage{amssymb}

\begin{document}
\begin{frontmatter}


\journal{SCES '04}


\title{Bound states in weakly disordered spin ladders}

%
%
%

\author[LP]{M. Arlego\corauthref{1}},
\author[BS]{W. Brenig},
\author[ST]{D.C. Cabra},
\author[BS]{F. Heidrich-Meisner},
\author[BS]{A. Honecker},
\author[LP]{G. Rossini}
%
%
\address[LP]{Departamento de F\'{\i}sica, Universidad Nacional de La Plata, Argentina}
\address[BS]{Institut f\"ur Theoretische Physik, Technische Universit\"at Braunschweig,  Germany}
\address[ST]{Laboratoire de Physique Th\'{e}orique, Universit\'{e} Louis Pasteur Strasbourg, France}



\corauth[1]{Corresponding Author: Departamento de F\'{\i}sica, Universidad Nacional de La Plata, C.C.\ 67 (1900) La Plata, 
Argentina. Phone: +54-221-424 6062. FAX: +54-221-425 2006, Email:\\ arlego@venus.fisica.unlp.edu.ar}


\begin{abstract}
We study the appearance of bound states in the spin gap of
spin-1/2  ladders induced by weak bond disorder. Starting
from the strong-coupling limit, i.e., the limit of weakly coupled dimers, 
we perform a projection 
on the single-triplet subspace and derive the position of 
bound states for the single impurity problem of one modified coupling
as well as for small impurity clusters. 
The case of a finite concentration of impurities is treated 
with the coherent-potential approximation (CPA) in the strong-coupling 
limit and compared with numerical results. 
Further,  we analyze the details in the structure of the density of states and relate their origin to the influence of  impurity clusters.    

\end{abstract}

%
%



\end{frontmatter}

\begin{figure}
     \centering
     \includegraphics[width=\columnwidth]{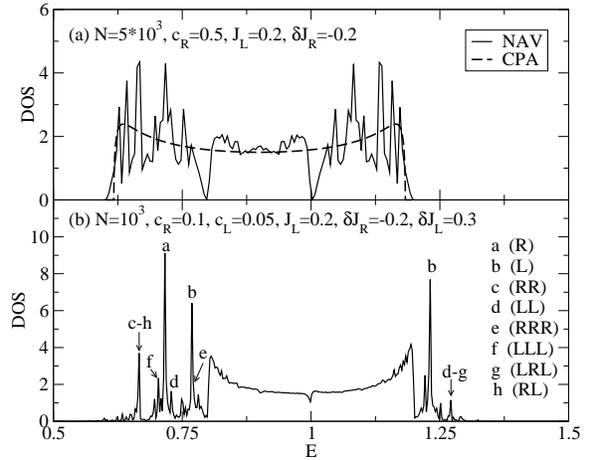}
     \caption{Panel (a): comparison of CPA and NAV for 
     a large concentration of rung impurities ($c_{\mathrm{R}}=0.5$, $c_{\mathrm{L}}=0$).
     Panel (b): NAV for bond-impurities on both rungs and legs. 
     The letters label peaks originating from certain impurity clusters as indicated in the legend.  See
     text for more details. In both cases, NAV was performed over typically 5000 random samples.}
     \label{fig:1} 
 \end{figure}

During the past decade spin ladder systems have attracted large
interest among theoretical and experimental physicists (see Ref.~\cite{dagotto99}
for a review) leading to considerable progress in understanding properties of 
such systems. In this paper we focus on spin ladders with weak bond-disorder 
and we allow  couplings along both rungs and legs to be modified. 
The Hamiltonian is 
\begin{equation}  
        H = \sum_{l=1}^{N}\lbrack J_{\mathrm{R},l}\vec{S}_{l,1}\hspace{-0.05cm}\cdot\hspace{-0.05cm} \vec{S}_{l,2}
	 + (J_{\mathrm{L},l,1}\vec{S}_{l,1}\hspace{-0.05cm}\cdot\hspace{-0.05cm}
              \vec{S}_{l+1,1}+ 1\leftrightarrow 2)\rbrack \,
	       .\label{eq:1} 
 \end{equation}
$N$ is the number of sites and $\vec{S}_{l,i}$ denotes a spin-$1/2$ operator acting on site
$l$ on leg $i$; $i=1,2$. 
As we have shown in Ref.~\cite{arlego04}
the presence of bond-impurities can induce bound-states in the spin gap. 
In extension of our previous work \cite{arlego04} we address two examples and present results for the density
of states  (DOS) of the single-triplet excitation: (i) a random, binary distribution of impurities 
on rung sites with a large impurity concentration and (ii)   binary distributions of modified couplings on legs and rungs. 
Thus, impurities are distributed according to the probability density
\begin{equation} P(J_{l}) = c_{\mathrm{R,L}} \delta(J_{l}-J'_{\mathrm{R,L}}) + (1-c_{\mathrm{R,L}})
\delta(J_{l}-J_{\mathrm{R,L}})\,
 \label{eq:3}
 \end{equation}
 where $J_{\mathrm{R,L}}$ denote the couplings of the pure system and $J'_{\mathrm{R,L}}$ 
 are the modified couplings.
In both cases, we assume the perturbations to be uncorrelated. Furthermore, the concentrations
for rung impurities $c_{\mathrm{R}}$ and for leg impurities $c_{\mathrm{L}}$ are  independent. 
\\\indent
To treat this problem, we proceed by first mapping the spin-$1/2$ operators
 onto bond-boson operators $s_l^{\dagger}$ and $t_{\alpha,l}^{\dagger}$ \cite{sachdev90}, 
 where  $s_l^{\dagger}$ creates a singlet on the $l$th rung and $t_{\alpha,l}^{\dagger}$
 a triplet excitation with orientation $\alpha=x,y,z$, respectively. Next, the singlet
 is integrated out and we restrict our study to the one-triplet subspace. 
 Thus, triplet interactions and quantum fluctuations are neglected which is 
 equivalent to first-order perturbation theory in $J_{\mathrm{L}}/J_{\mathrm{R}}$
 for the pure model.\\\indent 
 The resulting effective Hamiltonian
reads 
 \begin{equation} 
H_{\mathrm{eff}}=\sum_{l=1}^N \lbrack J_{\mathrm{R},l}t_{l}^{\dagger}t_{l}^{ } 
+  \frac{J_{\mathrm{L},l,1}+J_{\mathrm{L},l,2}}{4}
( t_{l+1}^{\dagger}t_{l}+\mbox{H.c.})\rbrack\, ;\label{eq:2}
\end{equation}
omitting the spin index $\alpha$ but keeping in mind that the spectrum is threefold degenerate.
The dispersion of the homogeneous model
is 
$\epsilon_k=J_{\mathrm{R}}+J_{\mathrm{L}} \cos(k)\,,$
$k$ being the momentum (see, e.g., Ref.~\cite{barnes93}).
The spectrum of this effective model in the presences of impurities can be obtained numerically for very large
system sizes and has to be sampled over many realizations of impurity-distributions
(numerical-impurity averaging, NAV). Analytically, the self-energy $\Sigma(E)$ ($E\to E+i0^+$) of the 
one-triplet Green's function $G(E)$ is accessible via diagrammatic techniques. Here we 
apply the coherent-potential approximation (CPA) to the case of $c_{\mathrm{R}}>0; c_{\mathrm{L}}=0$; the self-energy being a self-consistent
solution of the equation (see, e.g., Ref.~\cite{elliott74})
\begin{equation}
\Sigma(E) = \frac{c_{\mathrm{R}} \,\delta J_{\mathrm{R}}}{ 1 -  G(E)\, \lbrack \delta J_{\mathrm{R}}-\Sigma(E)  \rbrack }.
\,\label{eq:4}
\end{equation}
$\delta J_{\mathrm{R}}$ is defined as $J'_{\mathrm{R}}-J_{\mathrm{R}}$, compare Eq.~(\ref{eq:2}). \\
\indent 
Figure \ref{fig:1} contains  our results for the DOS.
First, we discuss the comparison of our numerical data and analytical computations for case (i), i.e.,
a binary distribution of rung-impurities with $J_{\mathrm{R}}=1,J_{\mathrm{L}}=0.2$, and
$\delta J_{\mathrm{R}}=-0.2$ as shown in Fig.~\ref{fig:1}~(a). 
The concentration of impurities is $c_{\mathrm{R}}=0.5$. Note that leg couplings 
are not perturbed. Both curves (solid: NAV, dashed: CPA) are symmetric with respect to 
$E=0.9$ while the original band for $c_{\mathrm{R}}=0$ was centered around $E=1$. 
In the limit
of $c_{\mathrm{R}}=1$ the band will be centered around $E=0.8$ possessing a band width of $0.2$. 
The CPA gives a qualitatively good description of the overall structure of the DOS.
As the  effective Hamiltonian (\ref{eq:2}) is that of a one-dimensional
non-interacting disordered system, one expects a smooth DOS in
the thermodynamic limit. However, the  NAV shows additional structures 
[see, e.g., the peaks a,\dots,h in Fig. 1 (b)] which are stable against variation
of $N$ for the system sizes investigated. These peaks can be attributed to the
presence of impurity clusters (see below).
 Small impurity clusters like, e.g., (RR) (two modified couplings on neighboring rungs)
or (R0R) (two modified couplings on  rungs separated by one non-perturbed rung) 
appear as soon as the concentration is finite and lead to discrete peaks in the DOS.
The corresponding eigen-energies can be computed exactly by solving Schr\"odinger's equation (see
Ref.~\cite{arlego04}).
\\
\indent
The influence of small impurity clusters will be further elucidated in the 
discussion of the second example (ii) where we have numerically computed spectra in the presence of finite concentrations
of both rung and leg impurities. Fig.~\ref{fig:1}~(b) shows the resulting DOS 
for $J_{\mathrm{R}}=1,J_{\mathrm{L}}=0.2,\delta J_{\mathrm{R}}=-0.2,\delta J_{\mathrm{L}}= J'_{\mathrm{L}}-J_{\mathrm{L}}=0.3 $ and $c_{\mathrm{R}}=0.1,c_{\mathrm{L}}=0.05$.
Note that $c_{\mathrm{L}}=1$ corresponds to all leg couplings on both legs being modified.\\
\indent 
In Fig.~\ref{fig:1}~(b), the shape of the original band is still visible for $0.8<E<1.2$ as the impurity
concentrations are fairly small. Both below and  above  the band, bound and anti-bound states 
appear. Their positions can be related to certain impurity clusters in the following way. 
We consider a single  impurity or   a single cluster in an otherwise clean system and compute
the eigen-energies of the corresponding states outside the original band, if such states are present
(see Ref.~\cite{arlego04} for a detailed discussion). Then, the main peaks visible in Fig.~\ref{fig:1}~(b)
can easily be  identified as being caused by small clusters or single impurities such as (R),  (RR),  or (RL).
Here, (RL) denotes a modified rung coupling and one of the neighboring leg-coupling modified.
\\
\indent Thus, at low concentrations, the structure 
of the DOS can basically be explained in terms of non-interfering, isolated impurity
clusters. The physical reason is that the wave-functions of bound-states are the less extended in
real space the larger the ratio $\delta J_{\mathrm{R}}/J_{\mathrm{R}}$, or $\delta
J_{\mathrm{L}}/J_{\mathrm{L}}$, respectively, is (see Ref.~\cite{arlego04} for  more details on 
the single-impurity wave-function).\\\indent
We acknowledge financial support by the Deutsche Forschungsgemeinschaft and the DAAD, Germany,
as well as 
Conicet, and 
Fundaci\'on Antorchas, Argentina.

%
%
%
%


\end{document}